\documentstyle[11pt]{article}
\input epsf.sty
\begin{document}
\parindent 1.3cm
\vspace*{-2cm}
\noindent
\hspace*{11cm}
UG--FT--60/96 \\
\hspace*{11cm}
hep-ph/9604287 \\
\hspace*{11cm}
March 1996 \\
\begin{center}
\begin{large}
{\bf Scale dependence of quark mass matrices \\
in models with flavor symmetries\\}
\end{large}
~\\
J. A. Aguilar--Saavedra, M. Masip \\
{\it Departamento de F\'{\i}sica Te\'{o}rica y del Cosmos \\
Universidad de Granada \\
18071 Granada, Spain}
\end{center}
\begin{abstract}
Numerical correlations between fermion 
masses and mixings could indicate the
presence of a flavor symmetry at high 
energies. In general, the search for these 
correlations using low-energy data requires 
an estimate of leading-log radiative corrections. We present 
a complete analysis of the evolution between 
the electroweak and the grand unification scales 
of quark mass parameters in minimal supersymmetric
 models. We take $M_t=180$ GeV and consider all possible values of 
$\tan \beta$. We also analize the possibility that the {\it top} and/or
the {\it bottom} Yukawa couplings result from an intermediate quasifixed
point (QFP) of the equations. We show that the quark mixings 
of the third family do {\it not} have a QFP behaviour
(in contrast to the masses, the renormalization of 
all the mixings is linear), and we evaluate 
the low-energy value of $V_{ub}$ which corresponds
to $V_{ub}(M_X)=0$. Then 
we focus on the renormalization-group corrections to {\it (i)} 
typical relations obtained in models with flavor 
symmetries at the unification scale and {\it (ii)} a
 superstring-motivated pattern of quark mass matrices.
We show that in most of the models the numerical prediction 
for $V_{ub}$ can be {\it corrected} in both directions (by
varying $\tan\beta$) due to  
{\it top} or {\it bottom} radiative corrections. 
\end{abstract}
\section{Introduction.}
The recent observation of the {\it top} quark, with 
a mass around $180$ GeV \cite{top95}, allows a more complete 
analysis of the Yukawa sector of the standard 
model. In particular, it allows an evaluation  
of the  running to higher  energy scales 
of the parameters in that sector. The perturbative 
unification of the gauge couplings obtained in minimal
supersymmetric (SUSY) scenarios 
suggests that the (nonsinglet) 
matter and gauge contents do not 
change up to energies around $10^{16}$ GeV. If that
 is the case, the first step to understand the flavor
 structure of the standard model is to evolve it up to 
those energies.
We present in the first part of this article an updated and complete 
analysis of the evolution from low energies to the 
unification scale $M_X$ of the 10 physical parameters 
in the quark mass matrices (6 masses, 3 mixing angles, 
and the complex CKM phase) in the minimal SUSY 
extension of the standard model (MSSM) \cite{olec91}.
We will study in detail
the behaviour of the mixing angles when the {\it top} and/or
{\it bottom} Yukawa couplings $h_{t,b}$ are large at $M_X$. 
In this regime \cite{hill81}
the renormalization-group corrections {\it focus}
any initial value of the coupling at $M_X$ to a
narrow interval $\delta h_t$ (around $h_t(M_Z)\approx 1.2$)
at low energy: ${(\delta h_t/ h_t)}(M_X)\gg
{(\delta h_t/ h_t)}(M_Z)$. We will analyze how this nonlinear
evolution of the couplings (quasifixed point (QFP) behaviour)
affects the mixings with the third generation; in
particular, we will find the low-energy value of the mixing 
$s_{13}$ (with $V_{ub}=s_{13} e^{-i\delta_{13}}$) that
corresponds to $s_{13}=0$ at $M_X$. 

On the other hand, the observed pattern of fermion 
masses and mixings does 
not look accidental, and suggests a symmetry in the 
Yukawa matrices as the origin of the hierarchies. 
Obviously, such a flavor symmetry should be formulated 
at the unification scale. Although the Yukawa matrices 
contain more free parameters than physical observables, 
it is not easy to find simple structures that are able 
to accommodate without need of fine tuning
the measured values of masses and mixings. 
As a matter of fact, the authors in Ref.~\cite{ramo93} 
classify the symmetric quark mass matrices with a maximal 
number of texture zeroes, and find that only five 
textures are acceptable experimentally. 
These matrices predict correlations between 
mass parameters that can be expressed in a simple
(approximate) form; we will analyze how the correlations 
run from $M_X$ to low energies. We will also analyze 
a particular scenario \cite{paco95}
derived from the heterotic string
which has been recently proposed as the only realistic
possibility among the models within its class.  
In this scenario (and also in the model proposed in  
\cite{dimo92}, with one more texture 
zero than the cases in \cite{ramo93})
the renormalization-group
corrections could be essential to obtain a predicted 
mixing $V_{ub}$ within the experimental limits. 

\section{Evolution of quark masses and mixings.}
The quark Yukawa sector of the MSSM can be expressed 
in terms of the superpotential
\begin{equation}
P=h^u_{ij}\; H Q_i u_j^c + h^d_{ij}\; H' Q_i d_j^c
\end{equation}
with $H\equiv (H^+\; H^0)$, $H'\equiv (H'^0\; H'^-)$
and $Q_i\equiv (u_i\; d_i)$.
In that sector there are 10 independent physical parameters: 
once the Yukawa matrices are diagonalized, only the
eigenvalues (3 in the up and 3 in the down quark matrices) 
and the Cabibbo-Kobayashi-Maskawa 
(CKM) matrix (with 3 mixing angles and a complex
 phase) appear in the Lagrangian. The procedure to obtain
 these  parameters at $M_X$ from the pole masses and the 
low-energy mixings is the following. For the heavier
 quarks, the perturbative pole mass $M_i$ is related
 to the running mass 
$m_i(M_i)$ in the $\overline{\mathrm MS}$ scheme by a simple
 expression \cite{gray90} (this change of scheme 
is numerically important due to the large size of $\alpha_s$). 
Taking $M_t=180$ GeV and $M_b=4.7$ GeV we obtain
\begin{equation}
m_b (M_b) = 0.880\; M_b  ~,~~ m_t (M_t) = 0.946\; M_t 
\end{equation}
The running masses of the lighter quarks are given 
at 1 GeV \cite{pdb94}:
$m_u = 0.0056\; {\rm GeV}$; $m_d =  0.0099\; {\rm GeV}$;
$m_s =  0.199\; {\rm GeV}$; 
$m_c =  1.35\; {\rm GeV}$. The masses evolve up to 
the lightest Higgs scale 
$m_h\approx M_Z$ due to gauge interactions only; this 
is a factor $0.58$ for the values at 1 GeV and $0.75$ 
for $m_b(M_b)$. It will be also convenient to run the 
{\it top} quark mass down to that scale;  
we obtain $m_t(M_Z)=1.05\;m_t(M_t)$. 

\begin{figure}[htb]
\begin{center}
\setlength{\unitlength}{1cm}
\begin{picture}(10,8.5)
\epsfysize=20cm
\epsfxsize=15cm
\put(-2.6,-7.5){\epsfbox{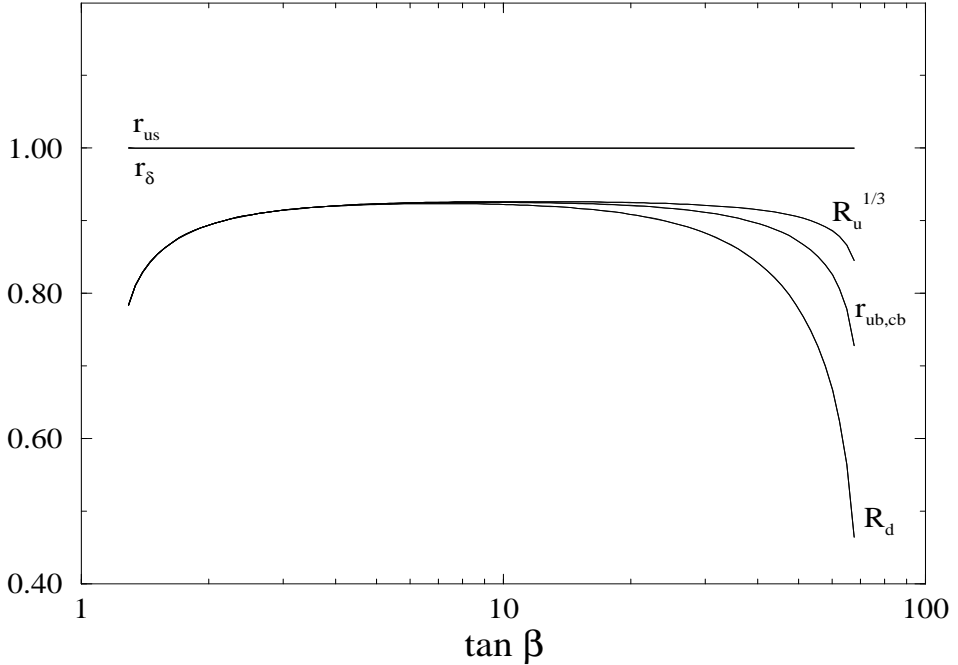} }
\end{picture}
\caption{Ratios $r_{ij}=V_{ij}(M_X)/V_{ij}(M_Z)$ 
($ij=us,cb,ub$), $r_{\delta}=\arg V_{ub}(M_X)/
\arg V_{ub}(M_Z)$, $R_u= {m_{u,c}(M_X)/m_t(M_X)\over 
m_{u,c}(M_Z)/m_t(M_Z)}$ and $R_d\equiv {m_{d,s}(M_X)/m_b(M_X)
\over m_{d,s}(M_Z)/m_b(M_Z)}$ for different values of 
$\tan\beta$.}
\label{gr:angulos}
\end{center}
\end{figure}

At $M_Z$ we find the Yukawa couplings to the lightest neutral
Higgs $h$, which correspond to the Yukawas in the 
standard model (we assume  $m_h=M_Z\ll m_{susy}= 250$ GeV)
\begin{equation}
\tilde h^u_{ij} = {m_{u_i}\over 174~{\rm GeV}} \delta_{ij} ~,~~
\tilde h^d_{ij} = {m_{d_i}\over 174~{\rm GeV}} V_{ij}^{*} 
\end{equation}
being $V_{ij}$ the CKM matrix at $M_Z$. From $M_Z$ to 
$M_t$ these Yukawas evolve due to the gauge and the 
(much smaller) Yukawa interactions of the light 
fermions; between $M_t$ and $m_{susy}$ the {\it top} 
Yukawa corrections are also important. From 
$m_{susy}$ up to $M_X\approx 10^{16}$ GeV we 
include the interactions of the SUSY particles 
and the extra Higgs scalars (the standard model and 
MSSM renormalization group equations can be 
found in Refs.~\cite{aras92,cast94}
respectively). The Yukawa couplings to the 
two Higgs doublets are (at $m_{susy}$)
\begin{equation}
h^u_{ij} = {\tilde h^u_{ij} \over \sin \beta} ~,~~
h^d_{ij} = {\tilde h^d_{ij} \over \cos \beta}
\end{equation}
where $\tan \beta$ is the ratio of VEVs giving 
mass to {\it down} and {\it up} quarks. At $M_X$ we 
diagonalize the Yukawa matrices (we express 
the eigenvalues as $h_i$, with $i=u,d...$) 
and find the CKM matrix. We will neglect corrections 
to mixing angles and light quark masses \cite{arka95}
proportional to the {\it soft} SUSY breaking 
parameters, although these corrections can be 
significant in the large $\tan\beta$ regime \cite{blaz95}. 
\begin{figure}[htb]
\begin{center}
\setlength{\unitlength}{1cm}
\begin{picture}(10,8.5)
\epsfysize=20cm
\epsfxsize=15cm
\put(-2.6,-7.5){\epsfbox{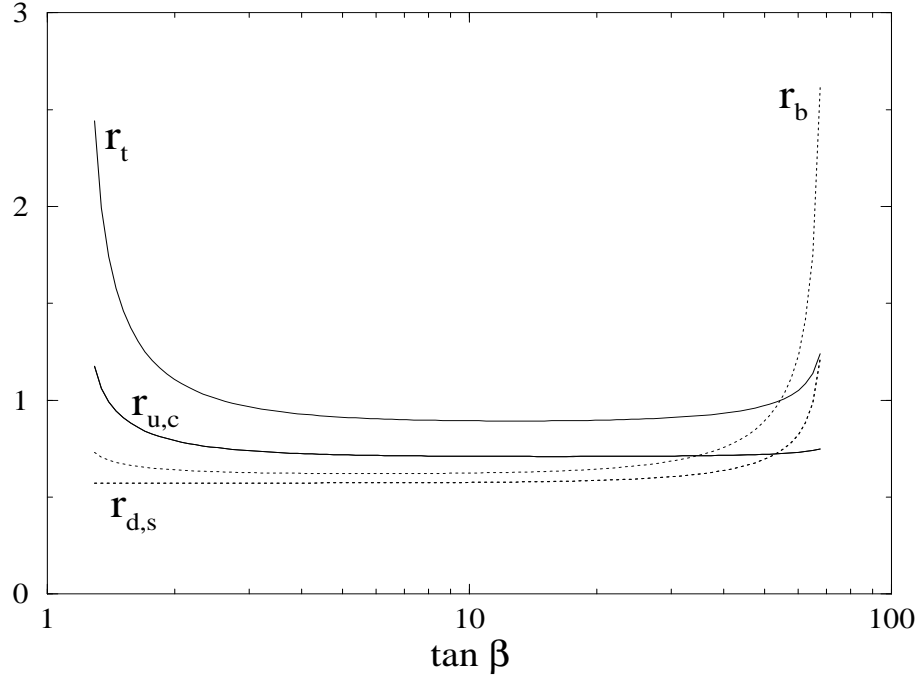} }
\end{picture}
\caption{Ratios $r_i=m_i(M_X)/m_i(M_Z)$ ($i=u,c,t,d,s,b$) 
for different values of $\tan\beta$.}
\label{gr:masas}
\end{center}
\end{figure}
We present in Figs.\ \ref{gr:masas}, \ref{gr:angulos}  
the evolution of masses and mixings for values of 
$\tan \beta$ between $1.29$ and $67.7$, which correspond
 respectively to 
$h_t/4\pi$ and $h_b/4\pi$ equal to $0.25$ at $M_X$ 
(we analyze below in detail the quasifixed point regions). 
In Fig.\ \ref{gr:masas} we plot 
$r_i\equiv {m_i(M_X)/ m_i(M_Z)}$ 
for the six quarks (note that the ratios of masses and 
Yukawas coincide), whereas Fig.\ \ref{gr:angulos} expresses the
ratios  $r_{ij}\equiv {V_{ij}(M_X)/ V_{ij}(M_Z)}$ 
($ij=us, cb, ub$) and $r_\delta\equiv {\arg [V_{ub}(M_X)]/
\arg [V_{ub}(M_Z)]}$ (in the Maiani parametrization). 
In order to compare the relative renormalization of
the mixings and 
different ratios of masses, we also plot $R_u^{1/3}$, $R_d$ in 
Fig.\ \ref{gr:angulos}, where
$R_u\equiv {m_{u,c}(M_X)/m_t(M_X)\over m_{u,c}(M_Z)/m_t(M_Z)}$
and  
$R_d\equiv {m_{d,s}(M_X)/m_b(M_X)\over m_{d,s}(M_Z)/m_b(M_Z)}$.
We take all the masses and mixings at $M_Z$ in their 
central value ($V_{us}= 0.221\pm 0.003$, $V_{cb}
= 0.040\pm 0.008$ and $|V_{ub}|= 0.0035\pm 0.0015$), 
with $M_t=180$ GeV and $\arg(V_{ub})=-\pi/2$.
From the analysis it follows that:
\begin{itemize}
\item The evolution of $m_u$ and $m_c$ coincide at 
the 0.04\%; the same happens for $m_d$ and $m_s$ 
(0.06\%) and $|V_{ub}|$ and $V_{cb}$ (0.05\%)

\item The running of the Cabibbo mixing $V_{us}$ 
and of the CKM phase (in $V_{ub}$) are smaller 
than the 0.03\% and 0.07\%, respectively. The evolution 
of masses and mixings is insensitive to the value 
of the complex phase.

\item The approximation \cite{ramo93}
\begin{eqnarray}
\frac{V_{ij}(M_X)}{V_{ij}(M_Z)} & = & \chi ~~ (~ij=
us,~cb,~ub~) \\
\frac{m_{u,c}(M_X)/m_t(M_X)}{m_{u,c}(M_Z)/m_t(M_Z)} 
& = & \chi^3 \\
\frac{m_{d,s}(M_X)/m_b(M_X)}{m_{d,s}(M_Z)/m_b(M_Z)} 
& = & \chi 
\end{eqnarray}
is excellent (error smaller than 1\%) for $\tan\beta 
\le 5$.
These expressions, with $\chi=(M_X/M_Z)^{(h_t/4\pi)^2}$,
are obtained assuming that the {\it top} 
Yukawa coupling is dominant and constant between $M_Z$
and $M_X$.
\end{itemize}
\begin{table}[htb]
\begin{center}
\begin{tabular}{cr@{--}l}
\hline
\hline
$V_{us}(M_Z)$ & 0.218 & 0.224 \\
$\sqrt{m_u/m_c}(M_Z)$ & \multicolumn{2}{c}{0.0609} \\
$\sqrt{m_d/m_s}(M_Z)$ & \multicolumn{2}{c}{0.224} \\
\hline
$V_{ub}(M_Z)$ & 0.002 & 0.005 \\
$\sqrt{m_u/m_t}(M_Z)$ & \multicolumn{2}{c}{0.0040} \\
$\sqrt{m_d/m_b}(M_Z)$ & \multicolumn{2}{c}{0.043} \\
\hline
$V_{cb}(M_Z)$ & 0.032 & 0.048 \\
$\sqrt{m_c/m_t}(M_Z)$ & \multicolumn{2}{c}{0.066} \\
$\sqrt{m_s/m_b}(M_Z)$ & \multicolumn{2}{c}{0.193} \\
\hline
\hline
\end{tabular}
\caption{Experimental value at $M_Z$ of quark 
mass ratios (the central value) and CKM
mixings (lower and upper limits). \label{tab:datos:iniciales} }
\end{center}
\end{table}

As showed in \cite{hill81}, the low-energy value of a 
large Yukawa coupling could be related to an intermediate 
QFP (previous to the Pendleton-Ross infrared fixed point 
\cite{pend81})
of the renormalization-group
equations. The effect of the QFP would be to {\it 
focus} any large initial value (at $M_X$) of the 
coupling to a narrow region at $M_Z$. In the MSSM, 
the {\it top} coupling approaches a QFP value for 
$\tan \beta\approx 1$, and the {\it bottom} coupling may 
approach a QFP value for large $\tan \beta$. For a 
{\it top} mass on its upper experimental limit, {\it both}
couplings could be related to QFPs \cite{frog93}. In 
Tables \ref{tab:datos:iniciales}--\ref{tab:fijo2}  
we show in some detail the behaviour of quark masses
and mixings near the QFP. 
Table \ref{tab:datos:iniciales} expresses the initial values (at $M_Z$) 
of masses and mixings, whereas in Table \ref{tab:fijo1} we write the values 
at $M_X$ for different values of $\tan\beta$ (which
correspond to $h_t(M_X)=(2,\;8)$ [{\it i.e.,} 
${h_t(M_X)/ 4\pi}=(0.16,0.64)$] 
and $h_b(M_X)=(2,\;8)$) for $M_t=180$ GeV. In 
Table \ref{tab:fijo2} both couplings are equal to 2 
and/or 8 at $M_X$ (this implies $M_t=190-220$ GeV and 
$\tan \beta= 60-69$). In all cases, the large Yukawa 
coupling at $M_Z$
varies very little around 1.2: 
${(\delta h_t/h_t)}(M_Z)\approx 0.06$ for 
any value of ${h_t/
4\pi}(M_X)$ larger than 0.32.
\begin{table}[htb]
\begin{center}
\begin{tabular}{cr@{--}lr@{--}lr@{--}lr@{--}l}
\hline
\hline
$\tan \beta$ & \multicolumn{2}{c}{1.22} & \multicolumn{2}{c}{1.44} & \multicolumn{2}{c}{65.0} & \multicolumn{2}{c}{69.5} \\
$h_t(M_Z)$ & \multicolumn{2}{c}{1.33} & \multicolumn{2}{c}{1.25} & \multicolumn{2}{c}{1.03} & \multicolumn{2}{c}{1.03} \\
$h_b(M_Z)$ & \multicolumn{2}{c}{0.028} & \multicolumn{2}{c}{0.031} & \multicolumn{2}{c}{1.16} & \multicolumn{2}{c}{1.24} \\
$h_t(M_X)$ & \multicolumn{2}{c}{8} & \multicolumn{2}{c}{2} & \multicolumn{2}{c}{1.17} & \multicolumn{2}{c}{1.51} \\
$h_b(M_X)$ & \multicolumn{2}{c}{0.024} & \multicolumn{2}{c}{0.021} & \multicolumn{2}{c}{2} & \multicolumn{2}{c}{8} \\
\hline
$V_{us}$ & 0.218 & 0.224 & 0.218 & 0.224 & 0.218 & 0.224 & 
0.218 & 0.224 \\
$\sqrt{m_u/m_c}$ & \multicolumn{2}{c}{0.0609} & \multicolumn{2}{c}{0.0609} & 
\multicolumn{2}{c}{0.0609} & \multicolumn{2}{c}{0.0608} \\
$\sqrt{m_d/m_s}$ & \multicolumn{2}{c}{0.224} & \multicolumn{2}{c}{0.224} & \multicolumn{2}{c}{0.224} & \multicolumn{2}{c}{0.224} \\
\hline
$V_{ub}$ & 0.00135 & 0.00337  & 0.00178 & 0.0042 & 
0.00156 & 0.00390 & 0.00125 & 0.00313\\
$\sqrt{m_u/m_t}$ & \multicolumn{2}{c}{0.00223} & \multicolumn{2}{c}{0.00310} & 
\multicolumn{2}{c}{0.00325} & \multicolumn{2}{c}{0.00289} \\
$\sqrt{m_d/m_b}$ & \multicolumn{2}{c}{0.0355} & \multicolumn{2}{c}{0.0396} & 
\multicolumn{2}{c}{0.0325} & \multicolumn{2}{c}{0.0236} \\
\hline
$V_{cb}$ & 0.0216 & 0.0324 & 0.0269 & 0.0403 & 
0.0249 & 0.0374 & 0.0200 & 0.0301 \\
$\sqrt{m_c/m_t}$ & \multicolumn{2}{c}{0.0366} & \multicolumn{2}{c}{0.0509} & 
\multicolumn{2}{c}{0.0533} & \multicolumn{2}{c}{0.0474} \\
$\sqrt{m_s/m_b}$ & \multicolumn{2}{c}{0.159} & \multicolumn{2}{c}{0.177} & \multicolumn{2}{c}{0.146} & \multicolumn{2}{c}{0.106} \\
\hline
$V_{ub}(M_Z)$ & \multicolumn{2}{c}{6.56 $10^{-6}$} & \multicolumn{2}{c}{3.54 $10^{-6}$} & \multicolumn{2}{c}{1.37 $10^{-6}$} & \multicolumn{2}{c}{1.28
$10^{-6}$} \\
\hline
\hline
\end{tabular}
\caption{Value at $M_X$ of quark mass ratios and 
CKM mixings for values of $\tan\beta$ which
correspond to large $h_t$ or $h_b$ at $M_X$ ($M_t=180$ GeV).
In the last line we write the value of $V_{ub}(M_Z)$
which would correspond to $V_{ub}(M_X)=0$. \label{tab:fijo1} }
\end{center}
\end{table}

The evolution of the mixing angles for Yukawa couplings 
near the QFP value presents the following features:
\begin{itemize}
\item The running of the 
mixings with the third family (and also the Cabibbo mixing)
is highly linear, in the sense that ${(\delta V_{ij} / 
V_{ij}) (M_X)}={(\delta V_{ij} / V_{ij})(M_Z)}$,
with ${ij}=us,ub,cb$. This fact means that the angles do 
not have a QFP behaviour. For example, 
for $h_t(M_X)=8$ the interval $V_{ub}(M_Z)=0.002$--$0.005$ 
evolves to $V_{ub}(M_X)=0.00135$--$0.00337$.

\item The evolution of $V_{us}$ is still smaller than 
 0.04\%. The difference 
between the running of $V_{cb}$ and $V_{ub}$ and of 
 the {\it up} and {\it charm} ({\it down} and {\it strange}) Yukawas, 
smaller than 0.02\% and 0.04\% (0.03\%) respectively, 
do not grow when increasing the Yukawa couplings at $M_X$.

\item The mixings with the third family $V_{ib}$ and
the ratios $\sqrt{m_{u,c}/m_t}$ and $\sqrt{m_{d,s}/m_b}$ 
tend to zero at $M_X$ when $h_t(M_X)$ and/or $h_b(M_X)$ 
increase. However, when increasing for example $h_t$, 
the three quantities decrease at different rate. As a 
consequence, it will be always possible to {\it correct}
a relation between masses and mixings   
by varying $\tan\beta$ and going to the 
adecuate ({\it top} or {\it bottom}) QFP region (see next section). 
\end{itemize}
\begin{table}[htb]
\begin{center}
\begin{tabular}{cr@{--}lr@{--}lr@{--}lr@{--}l}
\hline
\hline
$M_t$ & \multicolumn{2}{c}{203} & \multicolumn{2}{c}{220} & \multicolumn{2}{c}{193} & \multicolumn{2}{c}{214} \\
$\tan \beta$ & \multicolumn{2}{c}{63.5} & \multicolumn{2}{c}{59.8} & \multicolumn{2}{c}{68.8} & \multicolumn{2}{c}{66.8} \\
$h_t(M_Z)$ & \multicolumn{2}{c}{1.17} & \multicolumn{2}{c}{1.27} & \multicolumn{2}{c}{1.10} & \multicolumn{2}{c}{1.24} \\
$h_b(M_Z)$ & \multicolumn{2}{c}{1.13} & \multicolumn{2}{c}{1.07} & \multicolumn{2}{c}{1.23} & \multicolumn{2}{c}{1.19} \\
$h_t(M_X)$ & \multicolumn{2}{c}{2} & \multicolumn{2}{c}{8} & \multicolumn{2}{c}{2} & \multicolumn{2}{c}{8} \\
$h_b(M_X)$ & \multicolumn{2}{c}{2} & \multicolumn{2}{c}{2} & \multicolumn{2}{c}{8} & \multicolumn{2}{c}{8} \\
\hline
$V_{us}$ & 0.218 & 0.224 & 0.218 & 0.224 & 0.218 & 0.224 & 0.218 & 0.224 \\
$\sqrt{m_u/m_c}$ & \multicolumn{2}{c}{0.0609} & \multicolumn{2}{c}{0.0609} &  \multicolumn{2}{c}{0.0609} & \multicolumn{2}{c}{0.0609} \\
$\sqrt{m_d/m_s}$ & \multicolumn{2}{c}{0.224} & \multicolumn{2}{c}{0.224} &  \multicolumn{2}{c}{0.224} & \multicolumn{2}{c}{0.224} \\
\hline
$V_{ub}$ & 0.00146 & 0.0036 & 0.0012 & 0.00301 & 0.00121 & 0.00303 & 0.00101 & 0.00251 \\
$\sqrt{m_u/m_t}$ & \multicolumn{2}{c}{0.00276} & \multicolumn{2}{c}{0.00191} &  \multicolumn{2}{c}{0.00264} & \multicolumn{2}{c}{0.00183} \\
$\sqrt{m_d/m_b}$ & \multicolumn{2}{c}{0.0317} & \multicolumn{2}{c}{0.0295} &  \multicolumn{2}{c}{0.0233} & \multicolumn{2}{c}{0.0218} \\
\hline
$V_{cb}$ & 0.0234 & 0.0352 & 0.0192 & 0.0289 &  0.0194 & 0.0292 & 0.0161 & 0.0242 \\
$\sqrt{m_c/m_t}$ & \multicolumn{2}{c}{0.0453} & \multicolumn{2}{c}{0.0315} &  \multicolumn{2}{c}{0.0435} & \multicolumn{2}{c}{0.0301} \\
$\sqrt{m_s/m_b}$ & \multicolumn{2}{c}{0.142} & \multicolumn{2}{c}{0.132} &  \multicolumn{2}{c}{0.104} & \multicolumn{2}{c}{0.098} \\
\hline
\hline
\end{tabular}
\caption{Value at $M_X$ of quark mass ratios and 
CKM mixings for large $h_t$ and $h_b$ couplings at $M_X$
($M_t$ not fixed). \label{tab:fijo2} }
\end{center}
\end{table}

\begin{figure}[htb]
\begin{center}
\setlength{\unitlength}{1cm}
\begin{picture}(10,8.5)
\epsfysize=20cm
\epsfxsize=15cm
\put(-2.6,-7.5){\epsfbox{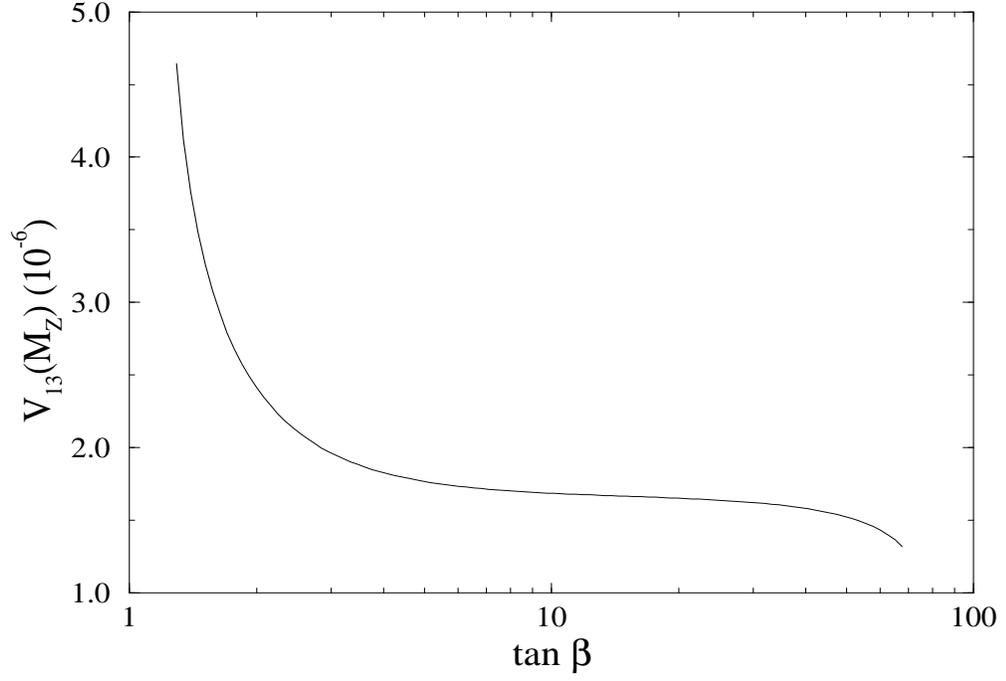} }
\end{picture}
\caption{Value of  $V_{ub}(M_Z)$ that would correspond to a zero value of 
$V_{ub}(M_X)$, plotted for different values of $\tan\beta$. The masses and the
mixings $V_{us}$, $V_{cb}$ are taken at their central value.\label{gr:v13}}
\end{center}
\end{figure}

To illustrate the size of the nonlinear effects 
(first point above), we will 
consider the case when $V_{ub}$ vanishes at $M_X$. 
The running from $M_X$ to $M_Z$ generates then a nonzero 
mixing. In Fig.\ \ref{gr:v13} we plot $V_{ub}(M_Z)$ 
for $\tan \beta$ between 1.29 and 67.7, with the 
fermion masses and the rest of mixings in their 
central value. We find a value much smaller than 
the experimentally preferred, although it grows 
when $\tan\beta$ decreases ({\it i.e.}, for a fixed
$m_t$, $h_t(M_X)$ 
increases). We also show in Table \ref{tab:fijo1} the 
low energy value of $V_{ub}$ which corresponds in each 
case to a vanishing mixing at $M_X$. Note that $V_{ub}$, 
the smallest mixing in the CKM matrix ($s_{13}$ in 
the Maiani parametrization), is a physical parameter 
whose zero value at one loop is not protected by  
any symmetry. A flavor structure giving $V_{ub}=0$
at $M_X$ would be caracterized by $V_{ub}(M_Z)\approx 10^{-5}$.

\section{Evolution of flavor relations}
We proceed now to study how the evolution from $M_X$
to $M_Z$ affects the relations between quark mass 
parameters obtained in models with flavor symmetries 
at the unification scale. First we will consider the 
5 symmetric
patterns with a maximal number of texture zeroes 
found in Ref.~\cite{ramo93}. These patterns depend 
on 7 complex parameters which, after phase redefinitions,
 are reduced to 7 moduli and 2 phases (3 phases 
in solution 2). The approximate analysis  
shows that it is possible to adjust the 
experimental masses and mixings without need of fine 
tuning. In particular, the absence of significant 
cancellations implies that the only role of the 2 complex
 phases in each pattern is to generate the CKM phase. 
As a consequence, the 7 moduli fit the 6 quark masses 
and 3 mixings giving two relations. One relation \cite{gatt68} is 
shared by all the cases (solutions 1 -- 5 in \cite{ramo93}):
\begin{equation}
V_{us}=\sqrt{m_d\over m_s}
\label{eq:rel1}
\end{equation}
(with complex corrections of modulus $\sqrt{m_u/ m_c}$ 
in solutions 1, 2, 4, 5). The second relation is
\begin{equation}
{V_{ub}\over V_{cb}}= \sqrt{m_u\over m_c}
\label{eq:rel2}
\end{equation}
for solutions 1, 2, 4 and
\begin{equation}
V_{ub}= \sqrt{m_u\over m_t}
\label{eq:rel3}
\end{equation}
for solutions 3 and 5 (in the last case there are 
complex corrections of order 
${m_t\over 2 m_c}V_{cb}^2\approx 20\%$). The masses 
and the rest of the mixings can be adjusted to their
central values, with an arbitrary CKM phase.

\begin{figure}[htb]
\begin{center}
\setlength{\unitlength}{1cm}
\begin{picture}(10,8.5)
\epsfysize=20cm
\epsfxsize=15cm
\put(-2.6,-7.5){\epsfbox{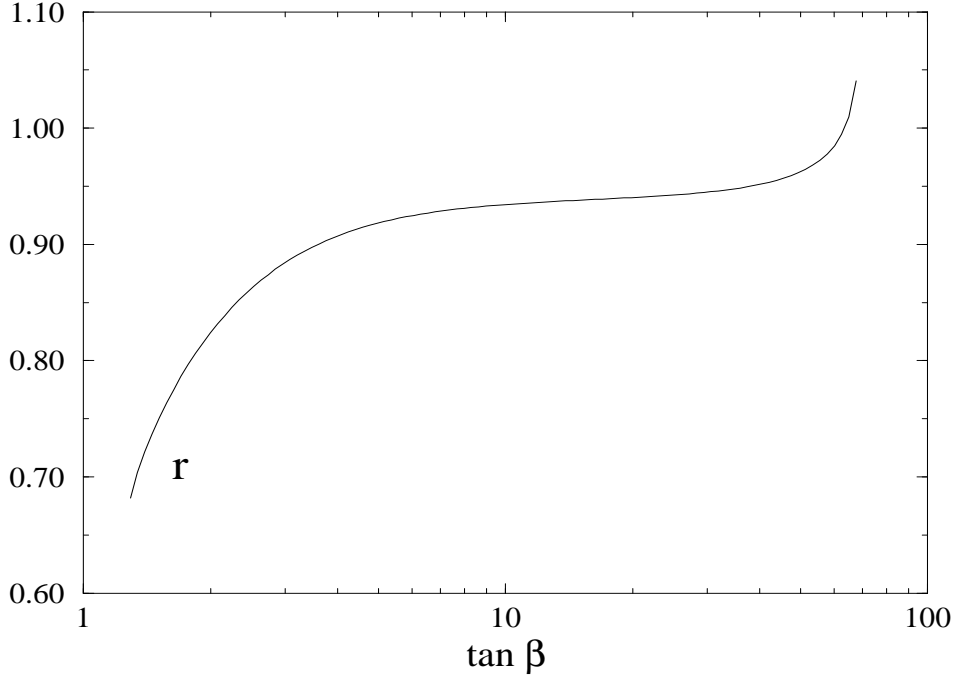} }
\end{picture}
\caption{Ratio $r$ defined in (\ref{ec:r}) for 
different values of  $\tan\beta$}
\label{gr:relacion:r}
\end{center}
\end{figure}

Relations (\ref{eq:rel1}--\ref{eq:rel3}) are 
stablished at $M_X$, and one has to
evolve the experimental quantities up to that 
scale in order to decide if they are acceptable. 
The running of the first two relations, however, 
is just a 0.2\% (smaller than corrections to the 
approximate diagonalization of the matrices). 
Taking the masses in their central values we have 
$\sqrt{m_d/ m_s}=0.223$ and
$\sqrt{m_u/ m_c}=0.064$, which compares well with 
the data  ($V_{12}=0.221 \pm 0.003$ and 
$V_{ub}/V_{cb}=0.08 \pm 0.02$ \cite{pdb94}). The relation 
$V_{ub}= \sqrt{m_u/ m_t}$ suffers sizeable 
renormalization-group corrections. At $M_Z$ we 
have $V_{ub}=0.0035\pm 0.0015$ and 
$\sqrt{m_u/m_t}\approx 0.76\sqrt{m_u(1\;{\rm GeV})/M_t}=
0.0040$. The running from $M_Z$ to $M_X$ can be 
expressed in terms of the ratio
\begin{equation}
r=\frac{\sqrt{m_u/m_t(M_X)}}{\sqrt{m_u/m_t(M_Z)}}  
\left/ \frac{V_{ub}(M_X)}
{V_{ub}(M_Z)} \right.
\label{ec:r}
\end{equation}
that we plot in Fig.\ \ref{gr:relacion:r} for different 
values of $\tan \beta$.
Note that for $\tan\beta < 62$, $V_{13}$ diminishes 
less than $\sqrt{m_u/m_t}$ ({\it i.e.,} $r<1$), while for larger 
values of  $\tan\beta $ we observe the opposite behaviour. 
If the masses and mixings were known with more accuracy, 
this fact could be used to {\it correct} the prediction
\begin{equation}
V_{ub}=r\; 0.0040
\end{equation}
in the preferred direction. 

The different running of the mixings and the ratios of quark   
masses involving the third family would also 
affect the symmetric texture with one more zero proposed 
in \cite{dimo92}. Those matrices predict 
$V_{cb}=\sqrt{m_c/m_t}$, a value that seems too large: 
 the values of the masses at $M_Z$ 
suggest $V_{cb}\approx 0.76\sqrt{m_c(1\;{\rm GeV})/M_t}=0.066$, 
while the experimental upper bound is 0.048. Since the 
evolution of $V_{ub}$ ($m_u$) and $V_{cb}$ ($m_c$) coincide, 
the running from $M_X$ will be simply expressed by the 
same factor $r$ in Fig.\ \ref{gr:relacion:r}: 
\begin{equation}
V_{ub}=r\; 0.066
\end{equation}
and the relation 
would be experimentally acceptable for $\tan \beta\le 1.4$
(with $M_t=180$ GeV).

We will finally analyze a pattern of quark matrices 
derived from the heterotic string. The matrices have 
been proposed \cite{paco95} as the only realistic possibility in a 
class of models compactified in the Tian-Yau manifold. 
Their structure is\footnote{We suppress an antisymmetric
entry proportional to the VEVs of an extra
Higgs doublet present in the model \cite{paco95}
since its presence would require a detailed analysis of
 flavor changing neutral currents.}
\begin{equation}
M_u=\left( \begin{array}{ccc}
0 & D_0 & C_0 \\ 
D_0 & 0 & B_0 \\ 
C_0 & B_0 & A_0
\end{array}
\right)~,~
M_d=\left( \begin{array}{ccc}
D'_0 & 0 & 0 \\
0 & C'_0 & B''_0 \\
0 & B'_0 & A'_0
\end{array}
\right)
\end{equation}
By a redefinition of the quark fields we can 
put these matrices in a more convenient form:
\begin{equation}
M_u=\left( \begin{array}{ccc}
0 & \tilde D & C \\ 
\tilde D & 0 & B \\ 
C & B & A
\end{array}
\right)~,~
M_d=\left( \begin{array}{ccc}
D' & 0 & 0 \\
0 & C' & 0 \\
0 & \tilde B' &  A'
\end{array}
\right)
\end{equation}
where $\tilde D$ and $\tilde B'$ are complex 
and the rest of the parameters are real and positive. 
As we will see, 
these 8 moduli and 2 phases can fit all the masses and 
the larger mixings to their central value and predict 
acceptable (but {\it large}) values for $V_{ub}$ and a 
nonzero (but {\it small}) complex CKM phase.  
The approximate diagonalization gives 
\begin{equation}
m_t= A ~,~~ m_c=\frac{B^2}{A} ~,~~ m_u= \left| 
\frac{2ABC\tilde D-A^2\tilde D^2}{A B^2} \right|
\end{equation}
\begin{equation}
m_b= A'~,~~ m_s = C'~,~~ m_d = D'
\end{equation}
and a CKM matrix (in the Maiani parametrization) with
\begin{eqnarray}
V_{us}\:\; & = & \left|\frac{BC-A\tilde D}{B^2} \right| \\
V_{cb}\;\; & = & \left| \frac{\tilde B'}{A'}-\frac{B}{A} \right| \\
\left| V_{ub}\right| & = & \left| \frac{(BC-A \tilde D) 
\tilde B'}{A B^2}-\frac{\tilde D}{B} \right|
\end{eqnarray}
In terms of physical quantities we have
\begin{equation}
V_{ub} =  \left( V_{us}V_{cb} + V_{12} \sqrt{\frac{m_c}{m_t}} 
e^{i\alpha} - \frac{m_u V_{cb}}{ 2 m_c V_{12}}
e^{i\beta} \right)
\end{equation}
where $\alpha$ and $\beta$ are independent complex 
phases (the dominant phase $\alpha$ is related 
to the phase of $\tilde B'$). At $M_Z$, 
for masses and mixings in their central values, the 
relation reads $V_{13}=0.0088 + 0.0146 e^{i\alpha}
-0.0004 e^{i\beta}\;.$ Then it seems 
that the small value of $V_{ub}$ requires a cancellation 
between the first two terms, with a best 
value $|V_{ub}|>0.0054$ (for $\alpha=\pi$). 
Renormalization-group corrections 
affect this relation due to the different running of 
$V_{cb}$ and $\sqrt{m_c/m_t}$, with the total effect 
captured again by the factor $r$ plotted in 
Fig.\ \ref{gr:relacion:r}
\begin{equation}
V_{ub}= 0.0088 + r\;0.0146 e^{i\alpha}
-0.00034 e^{i\beta}
\end{equation}
For low values of $r$, the lower bound for the 
predicted value of $|V_{ub}|$ decreases. For example, 
for $\tan\beta=1.5$ we have $|V_{ub}|>0.002$ and 
a CKM phase $\pi/2 \le \delta_{13} \le 3 \pi/2$, 
whereas $|V_{ub}|<0.005$ would imply $\tan\beta \le 30$ (for 
all the quark masses and
the rest of mixings in their central values). Note that 
for smaller values of $V_{cb}$, this bounds are relaxed.

\section{Conclusions}
The observed value of $M_t$ implies that $h_t$ is the
dominant term in the 
renormalization-group equations at large scales.
As a consequence, the corrections to the quark masses
lose universality and there appear nontrivial corrections
to the CKM mixings of the light quarks with the third family.
In addition, the low-energy value of $h_t$
could be related to a QFP of the equations: any {\it large}
value of $h_t(M_X)$ seems to converge to a narrow interval
around 1.2 at $M_Z$. In the MSSM with $M_t=180$ GeV this
forces a low value of $\tan\beta$, whereas an analogous 
situation occurs for $h_b$ in the large $\tan\beta$ regime.
For $M_t$ around 200 GeV and large $\tan\beta$, both 
low-energy Yukawa couplings would result from any large value 
of the couplings at $M_X$ (a large value of $\tan\beta$
could be also motivated by the possibility to relax the 
$R_b$ anomaly \cite{sola95}).
In this framework, we perform an
updated (with the new data for $M_t$) and complete (all values
of $\tan\beta$) analysis of the evolution from $M_Z$ and 
$M_X$ of all the physical observables in the quark Yukawa
sector of the MSSM. We study in detail the behaviour of the 
smallest CKM mixing $V_{ub}$ in the {\it top} and/or {\it bottom}
QFP regions, and we show that the evolution is linear: 
$\delta V_{ub}/V_{ub}(M_X) \approx \delta V_{ub}/V_{ub}(M_Z)$.
To illustrate the size the nonlinear corrections we analyze
the value of $V_{ub}(M_Z)$ which corresponds to 
$V_{ub}(M_X)=0$; we obtain $V_{ub}(M_Z)\le 10^{-5}$
(this value increases 
going to non-perturbative values of $h_t$ at 
$M_X$, {i.e.}, lowering $\tan\beta$).

Then we analize the renormalization-group corrections 
to fermion mass relations which appear in models with
flavor symmetries at $M_X$. In particular, we discuss
the relations obtained for symmetric mass
matrices with a maximal number of zeroes and in a 
superstring-motivated model. We show that in some 
relations the corrections
can be numerically important (they are essential in some of
the cases) and that they depend quite strongly on 
$\tan\beta$. In particular, for the relations analyzed the
corrections can be expressed in terms of the ratio $r$
in (\ref{ec:r}). We find remarkable that, for a fixed $M_t$, $r$ 
goes to zero decreasing $\tan\beta$ and grows ($r>1$) for
$\tan\beta$ large $(\tan\beta\geq 62)$. 
If the masses and mixings were measured with more accuracy, 
this fact could be used to conveniently
{\it correct} the relations by varying $\tan\beta$,
whereas if 
the Higgs sector of the MSSM were observed and $\tan\beta$ 
fixed, it could be used to exclude some of the quark 
mass matrix models.

\section{Acknowledgements}
This work was partially supported by CICYT under contract AEN94--0936,
by the Junta de Andaluc\'{\i}a and by the European Union under contract
CHRX--CT92--004. We thank F. del Aguila for useful comments.

\end{document}